%% file: lamt_final.tex
\documentclass[aps,prl,twocolumn,showpacs,superscriptaddress]{revtex4}

\usepackage{graphicx}
\usepackage{wrapfig}
\usepackage{amssymb}

\newcommand{\lpol}{P_{\text{n}}^{\Lambda}}

\newcommand{\lbarpol}{P_{\text{n}}^{\bar{\Lambda}}}

\begin{document}

%==========================================================================
% Title
%==========================================================================

\title{
  Transverse Polarization of $\Lambda$ and $\bar{\Lambda}$
  Hyperons in Quasireal Photoproduction
  \\ }

\include{authors}

%==========================================================================
% Abstract
%==========================================================================

\begin{abstract}

The HERMES experiment has measured the
transverse polarization of $\Lambda$  and $\bar\Lambda$ hyperons  
produced inclusively in quasireal photoproduction at 
a positron beam energy of 27.6~GeV\@.
The transverse polarization $\lpol$ of the $\Lambda$ hyperon is found to be
positive while the observed $\bar\Lambda$ polarization is compatible
with zero. The values averaged over the kinematic acceptance of HERMES are 
$\lpol = 0.078 \pm 0.006\,\text{(stat)} \pm 0.012\,\text{(syst)}$
and $\lbarpol = -0.025 \pm 0.015\,\text{(stat)} \pm 0.018\,\text{(syst)}$
for $\Lambda$ and $\bar\Lambda$, respectively. 
The dependences of $\lpol$ and $\lbarpol$ on  
the fraction $\zeta$ of the beam's light-cone momentum carried by the
hyperon and on the hyperon's 
transverse momentum $p_{\text{T}}$ were investigated.
The measured $\Lambda$ polarization 
rises linearly with $p_{\text{T}}$ and exhibits a different behavior
for low and high values of $\zeta$, which approximately correspond to 
the backward and forward regions in the center-of-mass frame 
of the $\gamma^*N$ reaction.

\end{abstract}

% insert suggested PACS numbers in braces on next line

% maketitle must be placed here
\maketitle

%==========================================================================
% Introduction
%==========================================================================

\section{Introduction}

In 1976, physicists at Fermilab measured the 
inclusive production of $\Lambda$ hyperons from high-energy proton-nucleon
scattering, and found a striking result: the $\Lambda$ particles produced 
in the forward direction and with transverse momenta greater than about 1 GeV
were highly polarized~\cite{Bunce1976}. 
Both the 300~GeV proton beam and the beryllium target were unpolarized. 
The $\Lambda$ polarization was transverse and negative, directed opposite 
to $ \hat{n} $, the unit vector along the direction 
$\vec{p}_{\text{beam}} \times \vec{p}_\Lambda $, which is normal to the 
production plane.
This ``self-polarization'' of final-state hadrons 
is observed quite commonly in the 
photoproduction of 
hyperons at low energies~\cite{Elsa,Clas}, and in exclusive 
reactions such as elastic $N N$ or $\pi N$ scattering~\cite{elastic}.
The $\Lambda$ polarization observable, proportional to 
$\vec{S}_\Lambda \cdot \hat{n}$,
where $\vec{S}_\Lambda$ is the spin vector of the $\Lambda$,
represents a single-spin asymmetry that is odd under naive time 
reversal. (Naive time reversal refers to the application of the time reversal 
operator $\hat{T}$ to each of the four-momenta in the reaction without
exchanging the initial and final states).
Given the $T$-even nature of the strong and electromagnetic interactions, 
such a naive $T$-odd observable must
arise through the interference of two $T$-even amplitudes: 
one that involves a helicity flip and one that does not~\cite{interference}. 
The surprise of the Fermilab result was
that the polarization also occured at high energies, in an inclusive 
measurement with many unobserved particles in the final state.
In this regime, perturbative QCD should accurately 
describe the partonic hard-scattering 
subprocess $a b \rightarrow c d$.
However, all helicity-flip amplitudes are
greatly suppressed in hard interactions as helicity is conserved
in the limit of massless quarks. 
The mechanism responsible for the polarization must thus arise from
the non-perturbative parts of the reaction, such as the fragmentation
process $c d \rightarrow \Lambda X$.
The production of a high-multiplicity final state at high energies
must involve a large number of amplitudes.
It seems remarkable that the phases of these amplitudes are
correlated to such a degree that a pronounced interference effect is observed.

The polarization of 
$\Lambda$ particles and other hyperons has now been observed 
and investigated
in many high-energy scattering experiments, with a wide variety of hadron 
beams and kinematic settings~\cite{Heller1996,Lach1996,WA89,HERA-B}.
The polarization of $\Lambda$ particles in particular is almost always 
found to 
be negative, as in the original $p N$ experiment.
A notable exception to this rule is the positive polarization
measured in $K^- p$~\cite{Gourlay86} and $\Sigma^- N$~\cite{WA89} 
interactions, %the only reaction measured to date
where the beam particles contain valence $s$ quarks.
A rather consistent kinematic behavior of the polarization has been observed: 
its magnitude increases
almost linearly with the transverse momentum $p_{\text{T}}$ of the 
$\Lambda$ hyperon up to a value
of about 1~GeV, where a plateau is reached.
The absolute polarization also rises with the Feynman variable $x_{\text{F}}$ 
with  values around
0.3 at $x_{\text{F}} \approx 0.7$. 

Possible mechanisms for the origin of this polarization were
reviewed in Refs.~\cite{Panagiotou1990} and~\cite{Soffer1999}
for example.
None of these models was able to account for the complete set 
of available measurements.
In particular, no model could explain the baffling 
pattern of anti-hyperon polarization. 
Anti-hyperons produced in $p N$ scattering contain no valence quarks in
common with the beam and are expected to have no polarization.
Zero polarization has indeed been consistently measured in the 
reaction $p N \rightarrow \bar\Lambda X$. However, studies of 
the reactions $p N \rightarrow \overline{\Xi}^- X$ and
$p p \rightarrow \overline{\Sigma}^+ X$ revealed anti-hyperon 
polarizations of the same sign and magnitude as those of the 
corresponding hyperons~\cite{AntiY}. These observations 
have presented a decade-long puzzle in non-perturbative QCD\@. To
our knowledge only one
possible solution has been suggested~\cite{Kubo} so far.

Given the large hyperon polarization observed in hadron-scattering experiments,
it is natural to wonder whether a non-vanishing polarization also occurs
in $\Lambda$ production by real and virtual photons at high energies. 
Very little experimental information exists about
this effect in photo- and electroproduction.
Transverse polarization in the inclusive photoproduction of neutral
strange particles  was investigated about 20 years ago at
CERN~\cite{Aston1982} and SLAC~\cite{Abe1984}. However,
the statistical accuracy of these data is limited.
The CERN measurements, 
for incident tagged photons with energies between 25 and 70~GeV, 
resulted in an average polarization of $0.06 \pm 0.04$. 
At SLAC, the overall polarization was observed to be $0.09 \pm 0.07$  
for  $\Lambda$ hyperons  produced using a 20~GeV photon beam. 
The SLAC experiment also investigated the dependence of 
the polarization on $x_{\text{F}}$ and observed a decrease, with the polarization
tending towards negative values  
for positive $x_{\text{F}}$.
%==========================================================================
% Experiment
%==========================================================================

\section{Experiment}

The HERMES experiment  offers an excellent opportunity 
to measure transverse $\Lambda$  and $\bar{\Lambda}$ 
polarization in the reaction
$\gamma^* N \rightarrow  \vec{\Lambda} X$,  using the 27.6~GeV positron beam 
of the HERA collider and an internal gas target. For simplicity the symbol 
$\Lambda$ will henceforward be used to refer to both the  $\Lambda$  and
$\bar{\Lambda}$ cases unless explicitely stated otherwise.
The HERMES detector~\cite{spectrometer} is a magnetic spectrometer whose
geometric acceptance is confined to two regions in scattering
angle, arranged symmetrically above and below the beam pipe. 
These regions are defined by the rectangular pole gaps in the spectrometer
magnet, and cover the ranges $\pm$(40--140)~mrad in the vertical component
of the scattering angle and $\pm$170~mrad in the horizontal component.
Between these regions is the horizontal septum plate of the magnet, 
which shields the HERA beams from the spectrometer's dipole field.
Thus, only particles produced with a polar angle greater than 40~mrad
with respect to the beam axis are visible. 
Since the standard HERMES trigger for deep-inelastic reactions requires an energy of 
more than 1.4~GeV or often even 3.5~GeV deposited in a lead-glass 
electromagnetic calorimeter,  
scattered positrons may be detected only for events with 
$Q^2$ above about 0.1~GeV$^2$ (where $-Q^2$ represents the 
four-momentum squared of the virtual photon).
In the study described in this paper, 
the detection of the scattered positron was not required and the final 
data sample is therefore dominated by the kinematic regime 
$Q^2 \approx 0$~GeV$^2$ 
of quasireal photoproduction where the cross section is largest. 
The scattered beam positron was detected
in coincidence with a $\Lambda$  in only 6\,\%
of the events.

A Monte Carlo simulation of the process using the PYTHIA event
generator~\cite{PYTHIA62}
and a GEANT~\cite{GEANT} model of the detector was used to estimate
the average kinematics of the $\Lambda$ sample.
An average virtual photon energy of $\langle \nu \rangle \approx 16$~GeV was obtained.
A total of about 70\,\% of the detected $\Lambda$ events
lie below $Q^2$ of 0.01~GeV$^2$, and about 90\,\% lie below 0.5~GeV$^2$.
Due to the long tail at higher values in the $Q^2$ distribution, 
the average $Q^2$ value is not representative of the typical event 
kinematics. 
The measurement is thus kinematically comparable to those at CERN and SLAC,
while offering a much higher statistical precision.
However, unlike in these two experiments, 
the kinematics of the quasireal photons are not known 
on an event-by-event basis. 

This analysis combines the 
data collected at HERMES in the years 1996 -- 2000.
The sample includes data taken with both longitudinally 
polarized and unpolarized targets, 
while the positron beam was always longitudinally polarized.
As the target spin direction was reversed every 90 seconds, 
the average target polarization was negligibly small.
The target species included hydrogen, deuterium, and a variety of unpolarized 
heavier gases.

%==========================================================================
% Extraction Method
%==========================================================================

\section{Extraction of the transverse polarization}

Because of the parity-conserving nature of the strong interaction,
any final-state hadron polarization in a reaction with unpolarized
beam and target must point along a pseudo-vector direction.
In the case of inclusive hyperon production, the only available direction
of this type is 
the normal $\hat{n}$ to the production plane formed by the cross-product
of the vectors along the laboratory-frame momenta of the 
positron beam ($\vec{p}_e$) and 
the $\Lambda$ ($\vec{p}_{\Lambda}$):
\begin{equation}
   \hat{n} = {\vec{p}_e \times \vec{p}_{\Lambda} \over |\vec{p}_e \times \vec{p}_{\Lambda} | }.
  \label{eq:nhat}
\end{equation}

\noindent By the same parity conservation argument, 
the polarization in this transverse (i.e., normal)
direction cannot depend 
linearly on the longitudinal polarization of the target ($P_{\text{T}}$) or the 
beam ($P_{\text{B}}$).
A dependence on their product $P_{\text{T}} P_{\text{B}}$, however, 
is not forbidden. 
In this analysis most of the $\Lambda$ data were collected using 
unpolarized targets, and the luminosity weighted value of
$P_{\text{B}} P_{\text{T}} $ was $0.0000\,\pm\,0.0005$ for
the entire data sample.

\begin{figure}[ht]
  \begin{center}
  \includegraphics[width=\columnwidth]{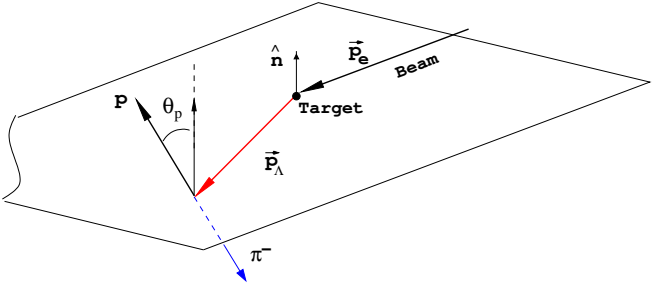}
  \caption{Schematic diagram of inclusive $\Lambda$ production and decay.
     The angle $\theta_p$ of the decay proton with respect to the
     normal $\hat{n}$ to the production plane is defined in the $\Lambda$
     rest frame.
  }
  \label{fig:diagram}
  \end{center}
\end{figure}

A kinematic diagram of inclusive $\Lambda$ production and the decay
$\Lambda \rightarrow p \pi^-$
is given in Fig.~\ref{fig:diagram}.
The $\Lambda$ decay is shown in the $\Lambda$ rest frame, where $\theta_p$ 
(see Eq.~\ref{eq:cmLam2})
is the angle of proton emission relative to the axis given by the normal
$\hat{n}$ to the scattering plane. Although $\hat{n}$ is defined in Eq.~\ref{eq:nhat} 
using vectors in the laboratory frame,
it is important to note that the direction 
is unaffected by the boost into the $\Lambda$ rest frame.  

The $\Lambda$ hyperon is a uniquely useful particle in spin physics:
the parity-violating nature of its weak decay 
$\Lambda \rightarrow p \pi^-$ results in an angular distribution where the 
protons are preferentially emitted along the spin direction of their parent $\Lambda$.
The angular distribution of the decay products of the $\Lambda$ may
thus be used to measure its polarization, providing 
a rare opportunity to explore spin degrees 
of freedom in the fragmentation process.
In the rest frame of the $\Lambda$ it has the form
\begin{equation}
  \frac{dN}{d\Omega_p} = {dN_0 \over d\Omega_p}
        ( 1 + \alpha \vec{P}^\Lambda \cdot \hat{k}_p ).
  \label{eq:cmLam}
\end{equation}
Here, $\hat{k}_p$ is the proton momentum unit vector
in the $\Lambda$ rest frame, 
$\vec{P}^\Lambda$ is the polarization of the $\Lambda$,
and $\alpha = 0.642\,\pm\,0.013$
is the analyzing power of the parity-violating weak decay  \cite{PDG}.
Assuming $CP$-invariance of the decay, the analyzing power
for the $\bar\Lambda$ is of
opposite sign ($\alpha^{\bar\Lambda} = -0.642$) 
~\cite{PDG}.
The quantity $dN_0 / d\Omega_p$ denotes the decay distribution
of \textit{unpolarized} $\Lambda$ particles. 
As described above, 
only the normal component $\lpol$ of the $\Lambda$ polarization may be non-zero
in the present analysis, and so Eq.~\ref{eq:cmLam} may be 
rewritten as 
\begin{equation}
  \frac{dN}{d\Omega_p} = {dN_0 \over d\Omega_p}
        ( 1 + \alpha \lpol \cos\theta_p ).
  \label{eq:cmLam2}
\end{equation}

For unpolarized $\Lambda$ particles the distribution of the decay
particles is isotropic and
$dN_0 / d\Omega_p$ is simply a normalization factor, 
independent of angle. In the case of limited spectrometer acceptance, however, 
it acquires a dependence on $\cos\theta_p$.

To extract the polarization of a sample of $\Lambda$ hyperons from 
the angular distribution of their decay products in the acceptance,
one may
determine the following moments:
\begin{equation}
  \langle \cos^m\theta_p \rangle \equiv \frac
        { \int \cos^m\theta_p \frac{dN}{d\Omega_p} d\Omega_p } 
        { \int              \frac{dN}{d\Omega_p} d\Omega_p } \equiv \frac
        { \int \cos^m\theta_p \frac{dN}{d\Omega_p} d\Omega_p } 
        { N^{\Lambda}_{\text{acc}} }, 
    \label{eq:cos}
\end{equation}
and
\begin{equation}
 \langle \cos^m\theta_p \rangle_0 \equiv \frac
        { \int \cos^m\theta_p \frac{dN_0}{d\Omega_p} d\Omega_p } 
        { \int              \frac{dN_0}{d\Omega_p} d\Omega_p } \equiv \frac
        { \int \cos^m\theta_p \frac{dN_0}{d\Omega_p} d\Omega_p } 
        { N_{0,\text{acc}}^{\Lambda} }, 
    \label{eq:coszero}
\end{equation}
where $m = 1,2, ..$~. The symbol $\langle...\rangle$ represents 
an average over an actual
data sample, while $\langle...\rangle_0$ denotes an average over a
hypothetical purely-unpolarized sample of  $\Lambda$ particles with an
isotropic decay distribution.
$N_{\text{acc}}^{\Lambda}$ and $N_{0,\text{acc}}^{\Lambda}$ are equal to the total number
of $\Lambda$ events for the same luminosity accepted by the spectrometer.
They are related by
\begin{equation}
N_{\text{acc}}^{\Lambda} = N_{0,\text{acc}}^{\Lambda}(1 + \alpha \lpol \langle \cos\theta_p 
\rangle_0).
    \label{eq:pol}
\end{equation}
Combining Eqs.~\ref{eq:cmLam2} - \ref{eq:pol} one obtains
\begin{equation}
  \langle \cos^m\theta_p \rangle = \frac
        { \langle \cos^m\theta_p \rangle_0 + \alpha \lpol
          \langle \cos^{m+1}\theta_p \rangle_0 }
        { 1 + \alpha \lpol \langle \cos\theta_p \rangle_0 }.
  \label{eq:main}
\end{equation}

The extraction of the $\Lambda$ polarization $\lpol$ from the 
experimental data is based on Eq.~\ref{eq:main}. The `polarized'
moments
$\langle \cos^m\theta_p \rangle$ can be determined by taking an 
average over the experimental data set: 
\begin{equation}
  \langle \cos^m\theta_p \rangle = \frac{1}{N_{\text{acc}}^\Lambda} 
        \sum \limits_{i=1}^{N_{\text{acc}}^\Lambda}\cos^m\theta_{p,i}.
  \label{eq:aver}
\end{equation}

The `unpolarized' moments $\langle \cos^m\theta_p \rangle_0$
cannot be extracted directly from the data as no sample of 
unpolarized $\Lambda$ hyperons is available. 
Fortuitously, however, the extraction of the transverse $\Lambda$ polarization 
from the HERMES data is greatly
simplified by the up/down mirror symmetry of the HERMES spectrometer, even in
the case of limited acceptance.
It can be readily shown that this geometric symmetry leads to the relation
\begin{equation}
  \langle \cos^m\theta_p \rangle_0^{top} =
        (-1)^m \langle \cos^m\theta_p\rangle_0^{bot}, 
 \label{eq:sym1}
\end{equation}
where $top$ and $bot$ specify 
events in which the hyperon's momentum was directed above or below the
midplane of the spectrometer.
Consequently all `unpolarized' uneven moments of the full acceptance
function ($top$ plus $bot$) are zero,
and all even `polarized' moments are equal to the `unpolarized' ones:
\begin{equation}
\langle \cos^m\theta_p \rangle = \langle \cos^m\theta_p \rangle_0
~~m = 2, 4, ... ~.
\label{eq:sym2}
\end{equation}

The first moment of $\cos\theta_p$
may be calculated separately for the {\it top} and {\it bot} data samples 
to account for a possible 
difference in the overall
efficiency of each detector half. 
Using the symmetry relations (Eqs.~\ref{eq:sym1} and 
\ref{eq:sym2}), one obtains from Eq.~\ref{eq:main} a system of two coupled
equations for $\alpha\lpol$ and $\langle \cos\theta_p \rangle_0^{top}$: 
\begin{equation}
\alpha\lpol = \frac {c_+/\langle \cos^2\theta_p \rangle} 
 {1 - \langle \cos\theta_p \rangle_0^{top}~c_-/\langle \cos^2\theta_p \rangle},
\end{equation}
\begin{equation}
\langle \cos\theta_p \rangle_0^{top} =\frac {c_-} {1 - c_+\alpha\lpol},
\end{equation}
where $2c_+$ ($2c_-$) is the sum (difference) of 
$\langle \cos\theta_p \rangle^{top}$ and $\langle \cos\theta_p \rangle^{bot}$.
This system of coupled equations can be solved iteratively. 
The iteration converges quickly. 
If one takes 
$\alpha\lpol = c_+/\langle \cos^2\theta_p \rangle$ and 
$\langle \cos\theta_p \rangle_0^{top} = c_-$ for the first iteration, then 
the solution of the second iteration for $\lpol$ and 
$\langle \cos\theta_p \rangle_0^{top}$ reads:
\begin{equation}
\alpha\lpol = \frac {c_+/\langle \cos^2\theta_p \rangle} 
 {1 - c_-^2/\langle \cos^2\theta_p \rangle},
  \label{eq:master}
\end{equation}
\begin{equation}
\langle \cos\theta_p \rangle_0^{top} =\frac {c_-} 
{1 - c_+^2/\langle \cos^2\theta_p \rangle}.
 \label{eq:cos0}
\end{equation}

Eq.~\ref{eq:master} 
was used to determine the results  presented 
in this paper.

The results for the `unpolarized' first moment of $\cos\theta_p$ 
determined in various kinematic bins from data were found to be in very good 
agreement with those
obtained from a Monte Carlo simulation of the detector.  

%==========================================================================
% Data Analysis
%==========================================================================

\section{Event selection}

The kinematics of the $\Lambda$  hyperons whose decay
products are both within the angular acceptance of the HERMES
spectrometer are such that the proton 
momentum is always much higher
than that of the pion.
These low-momentum pions are often bent so
severely in the spectrometer magnet that they fail to reach the
tracking chambers and particle identification detectors in the backward 
half of the spectrometer.
However, it is possible to evaluate the momentum of such ``short tracks''
using the hits recorded by the HERMES Magnet Chambers, a series of
proportional chambers located between the poles of the
spectrometer magnet~\cite{MagCh}.
The acceptance for $\Lambda$
hyperons can be increased by almost a factor
of two when these pion ``short tracks'' are included in the analysis.
As non-pions in coincidence with protons are rare, particle identification (PID) 
is not essential for these 
tracks.
In contrast, PID of the decay proton
is important for background reduction. 
For the data recorded prior to 1998, this was provided by a threshold
\v{C}erenkov counter~\cite{spectrometer}, which was then
replaced by a dual-radiator Ring-Imaging \v{C}erenkov detector (RICH)~\cite{RICH}. 
Proton candidates were therefore required to be a positive 
hadron with the highest-momentum (leading hadron) having a 
``long track'', i.e., a track that passed through
all detectors of the spectrometer, and to be not identified as a pion. 

$\Lambda$ events were identified through the
reconstruction of secondary vertices in events containing oppositely charged 
hadron pairs.
Two spatial vertices were reconstructed for each event.
First the secondary (decay) vertex was determined from the intersection
(i.e., point of closest approach) of the proton and 
pion tracks. 
The hyperon track was then reconstructed using this decay vertex
and the sum of the proton and pion 3-momenta. The intersection of  
this track with the
beam axis determined the primary (production) vertex.
For both vertices the distance of closest approach was required to be less
than 1.5~cm.
Only those events with the primary vertex inside the 40~cm long target cell
were selected. 
All tracks were also required to satisfy a series of fiducial-volume cuts
designed to avoid the 
edges of the detector. Furthermore the 
two hadron tracks were required to be reconstructed in the same spectrometer 
half to avoid effects caused by a possible misalignment 
of the two spectrometer halves relative to each other.

Hadrons
emitted from the primary vertex were suppressed by two vertex
separation requirements. The transverse distance between the decay vertex
and the beam axis was required to be larger than 1~cm. In the longitudinal
direction the requirement  $z_2 - z_1 > 15 (20)$~cm was imposed for $\Lambda$ 
candidates,
with $z_1$ and $z_2$ representing
the coordinates of the primary and secondary vertex positions along the
beam direction. The chosen values of this vertex separation requirement were a 
compromise between statistical precision and low background of the data sample.

\begin{figure}[ht]
  \begin{center}
  \includegraphics[width=\columnwidth]{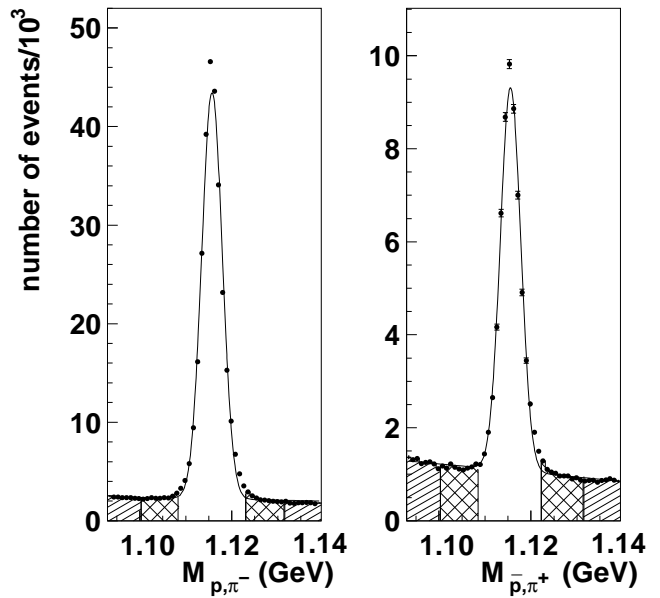}
  \caption{Invariant mass distributions for $\Lambda$ and $\bar \Lambda$    
        events. The central region was used for 
        the determination of the  $\Lambda$  ($\bar \Lambda$) polarization.
        The shaded areas indicate the invariant-mass intervals used for the
        determination of the background polarization.
  }
  \label{fig:mass}
  \end{center}
\end{figure}

The resulting $p \pi^-$ and $\bar p \pi^+$ invariant mass distributions 
are shown in Fig.~\ref{fig:mass}.
The fitted mean value for the $\Lambda$ ($\bar\Lambda$) mass is  
1.1157~GeV (1.1156~GeV)
with a width of 
$\sigma = 2.23$~MeV (2.20~MeV).
For the polarization analysis, $\Lambda$ and $\bar\Lambda$ events
within a $\pm 3.3\,\sigma$ invariant mass window around the mean value 
of the fitted peak
were chosen, and a background-subtraction procedure was applied
as described below.
The final data sample contained around 
$259 \times 10^3$ $\Lambda$ and $51 \times 10^3$ $\bar\Lambda$ events.

%

%==========================================================================
% Results
%==========================================================================

\section{Results}

The transverse polarization for the $\Lambda$ and $\bar\Lambda$
data samples was extracted using 
Eq.~\ref{eq:master}. 
The contribution of the background under the $\Lambda$ invariant mass peak
to the polarization 
was estimated using a side-band subtraction method. 
An independent polarization analysis was performed in 
each kinematic bin of interest. 
For each bin in $\zeta$ or $p_{\text{T}}$ (described below), 
the invariant mass 
spectrum was fit with a Gaussian plus a third-order polynomial. 
The fit was used to determine the number of signal and background 
events within a $\pm 3.3\,\sigma$ window around the peak.
The polarization was calculated for the events within
this central window, as well as within 
four ``sideband'' windows with widths of around 8~MeV,
two in the low- and two in the high-mass background regions, 
as indicated by the shaded 
areas in Fig.~\ref{fig:mass}. 
The polarizations extracted
from the sidebands were interpolated to obtain the background polarization
at the peak mass. The fraction of background events 
$\epsilon = \frac {N_{bgr}}{N_{\Lambda} + N_{bgr}}$ within the peak was
typically of order $15~\%$. The transverse polarization within the $\Lambda$
peak was corrected for this background contribution in each kinematic bin 
as follows

\begin{equation}
P_{\text{n}}^{\Lambda} = \frac {P_{\text{n}}^{\Lambda + bgr} - 
\epsilon P_{\text{n}}^{bgr}}{1 - \epsilon}.
  \label{eq:bgr}
\end{equation}

The interpolated background polarization $P_{\text{n}}^{bgr}$ was
around $0.12 \pm 0.01$ $(0.13 \pm 0.02)$ for the $\Lambda$ ($\bar{\Lambda}$) sample. 
Because of the small background contamination, the net correction 
to the $\Lambda$ and $\bar\Lambda$ polarization was on average below 0.01.
The results for the extracted value of the transverse 
$\Lambda $ polarization were stable within the statistical
uncertainty when the longitudinal vertex separation requirement was varied 
between 10 and 25~cm.

In order to estimate the systematic uncertainty of the measurement, 
similar analyses were carried out for 
reconstructed $h^+h^-$ hadron pairs, with leading  
positive hadrons ($\Lambda$-like case) and with leading negative hadrons 
($\bar\Lambda$-like case). No PID (apart from lepton rejection) was applied to these hadrons, and so the sample was likely dominated by $\pi^+ \pi^-$ 
pairs. Events within two mass windows 
above and below the $\Lambda$ mass window 
($ 1.093~\text{GeV} < M_{h^+h^-} < 1.108~\text{GeV}$, and 
$1.124~\text{GeV} < M_{h^+h^-} < 1.139~\text{GeV}$)
were selected, where $M_{h^+h^-}$ was determined by assuming for the 
leading/non-leading particles the proton/pion masses respectively. 
Instead of requiring a displaced decay vertex,  their point 
of closest approach
was required to be inside the target cell.
False polarization values of 
$0.012 \pm 0.002$ and 
$0.018 \pm 0.002$ were found in the 
$\Lambda$-like and $\bar\Lambda$-like cases, respectively.

As a second measure of the systematic uncertainty a sample of 
$K_s^0 \rightarrow \pi^+~\pi^- $ events was used. The long-lived  $K_s^0$
provides a similar event topology to the $\Lambda$ with two separated vertices.
The false  polarization of $K_s^0$  was found to be
$0.012 \pm 0.004$  in the $\Lambda$-like case (with a leading $\pi^+$) 
and $0.002 \pm 0.004$ in the $\bar\Lambda$-like case.  

Possible detector misalignments could lead to imperfections in the up/down 
symmetry of the spectrometer. In order to estimate the effect of such
misalignments on the measured polarizations, Monte Carlo simulations were
performed using a spectrometer description with the top and bottom halves
misaligned by $\pm 0.5$~mrad. Four samples were generated, with input
polarizations of 0, 0.05, 0.1 and 0.2, respectively. In addition a background 
polarization of 0.15 was included to better simulate the experimental
situation. The polarizations
extracted from these Monte Carlo data samples were in agreement with the input values 
within the statistical uncertainty of 0.005. A second potential
source of a top/bottom spectrometer 
asymmetry is trigger inefficiency. 
This was also investigated using Monte Carlo simulations. It was 
found that even an unrealistically large difference of 30\% in the
top/bottom efficiency resulted in the reconstructed polarization being consistent with
the generated one.

From the results of these studies the systematic uncertainties on the
$\Lambda$ and $\bar\Lambda$ transverse polarizations were taken to be
0.012 and 0.018, respectively.

\begin{figure}[ht]
  \includegraphics[width=7cm]{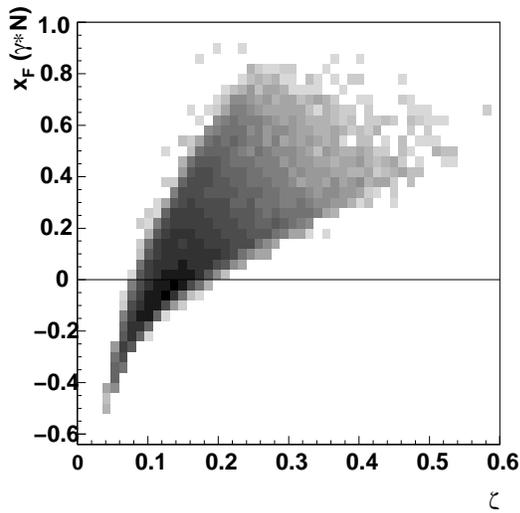} 
  \caption{Correlation between $x_{\text{F}}$, evaluated in the $\gamma^* N$
        system, and the light-cone fraction $\zeta$
        determined in the $eN$ system, as determined from a PYTHIA
        Monte Carlo simulation.}
  \label{fig:hemispheres1}
\end{figure}

\begin{figure}[ht]
  \includegraphics[width=7cm]{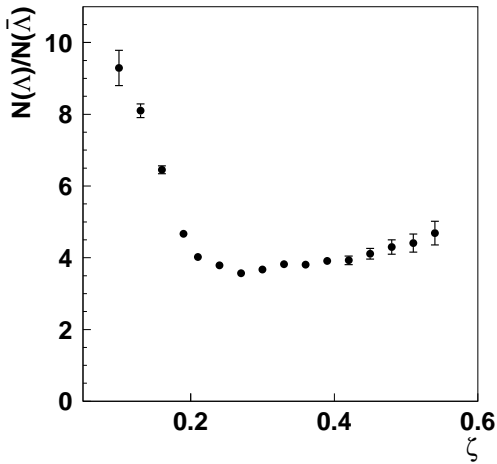} 
  \caption{Ratio of $\Lambda$ to $\bar\Lambda$ yields versus light-cone
        fraction $\zeta$ observed in the
        data, after background subtraction.
}
  
 \label{fig:hemispheres2}  
\end{figure}

The good statistical accuracy of the full inclusive data set allows 
the dependence of the $\Lambda$ and $\bar\Lambda$ polarization on 
certain kinematic variables to be studied.
As mentioned earlier, information on the virtual-photon kinematics 
is not known on an event-by-event basis; consequently,
only kinematic variables related to the $eN$ system are available.
However, one may analyze the data using
the kinematic variable $\zeta \equiv (E_\Lambda+p_{z\Lambda})/(E_e+p_e)$,
where $E_\Lambda$, $p_{z\Lambda}$ are the energy and 
  $z$-component of the $\Lambda$ momentum
  (where the $z$-axis is defined as the lepton beam direction), 
and $E_e$, $p_e$ are the energy and momentum of
the positron beam. This variable is the fraction of the
beam positron's light-cone momentum 
carried by the outgoing $\Lambda$.
It is an approximate measure of whether 
the hyperons were produced in the forward or backward region in the 
$\gamma^* N$ center-of-mass system. 
The natural variable to use to separate these kinematic regimes would
be the Feynman variable $x_{\text{F}} = p_\parallel^\Lambda/p_{\parallel \text{max}}^\Lambda$ 
evaluated in the $\gamma^* N$ system, where $ p_\parallel^\Lambda $ is the $\Lambda$'s
momentum along the virtual-photon 
direction, and $p_{\parallel \text{max}}^\Lambda$ is its
maximum possible value,
but this variable is not
available in an inclusive measurement. Nevertheless,
as shown in Fig.~\ref{fig:hemispheres1}, a simulation of the reaction
using the PYTHIA Monte Carlo
reveals a useful correlation 
between $\zeta$ 
and $x_{\text{F}}$.  
In particular, all events 
at $\zeta \ge 0.25$ are produced in the  
kinematic region $x_{\text{F}} > 0$, and
for $\zeta < 0.25$ there is a mixture of events originating 
from
the kinematic regions with $x_{\text{F}} > 0$ and $x_{\text{F}} < 0$. 
An indication that the dominant production mechanism changes at $\zeta$
values around 0.25 can be observed in the ratio of $\Lambda$ to $\bar\Lambda$ 
yields displayed in Fig.~\ref{fig:hemispheres2}.
The yields are not corrected for acceptance as PYTHIA Monte Carlo studies
indicate that the detection efficiencies for $\Lambda$ and $\bar\Lambda$
are the same.
Above $\zeta \approx 0.25$, an approximately 
constant ratio of about 4 is seen. At 
lower
values the ratio increases significantly, likely indicating the
influence of the nucleon target remnant in $\Lambda$ formation.


\begin{figure}[ht]
  \begin{center}
    \includegraphics[width=8cm]{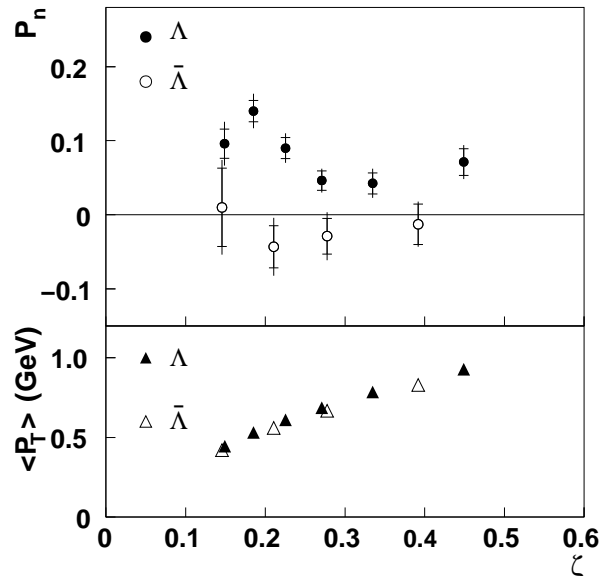}
  \caption{Transverse polarizations $\lpol$ and $\lbarpol$ (upper panel) 
        and mean $\langle p_{\text{T}} \rangle$ (lower panel)
        as functions of $\zeta=(E_\Lambda + p_{z\Lambda})/(E_e+p_e)$.
        The inner error bars represent the statistical 
        uncertainties, and the outer error bars represent the statistical and
        systematic uncertainties added in quadrature. 
  }
  \label{fig:zeta}   
  \end{center}
\end{figure}

The $\Lambda$ and $\bar\Lambda$ polarizations are shown 
as functions of $\zeta$ in Fig.~\ref{fig:zeta}.
The $\Lambda$ polarization is about 0.10 in the region 
$\zeta < 0.25$, and about 0.05 at higher $\zeta$.
Combining all kinematic points together, the average $\Lambda$
transverse polarization is
\begin{equation}
\lpol = 0.078 \pm 0.006 \text{(stat)} \pm 0.012 \text{(syst)}.
\end{equation}
  For the $\bar\Lambda$ measurement, no kinematic dependence is observed
  within the statistical uncertainties. 
The net $\bar\Lambda$ transverse polarization is
  \begin{equation}
    P_{\text{n}}^{\bar\Lambda} = -0.025 \pm 0.015 \text{(stat)} \pm 0.018 \text{(syst)} . 
  \end{equation}

\begin{figure}[ht]
\includegraphics[width=\columnwidth]{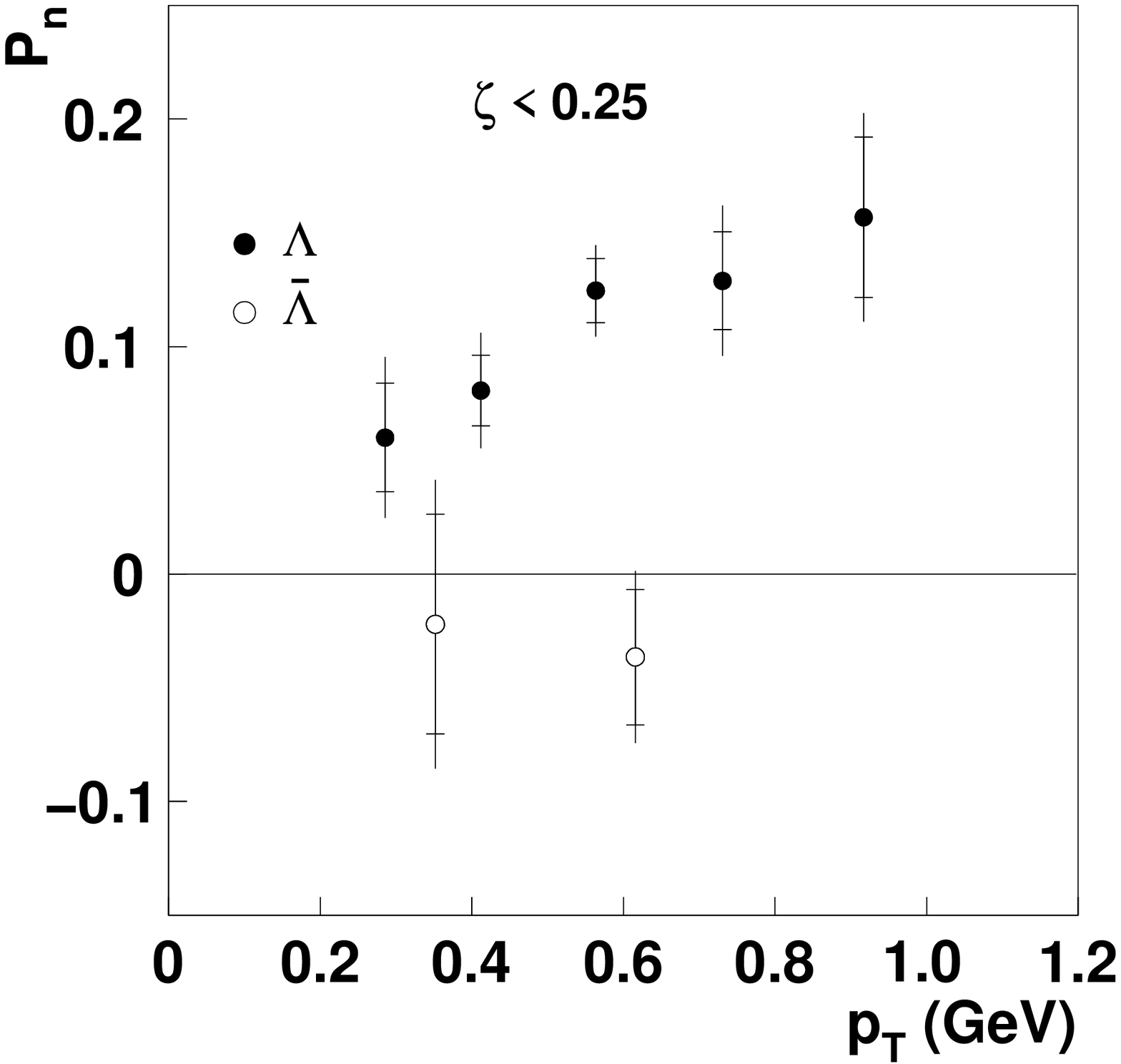} \\
\includegraphics[width=\columnwidth]{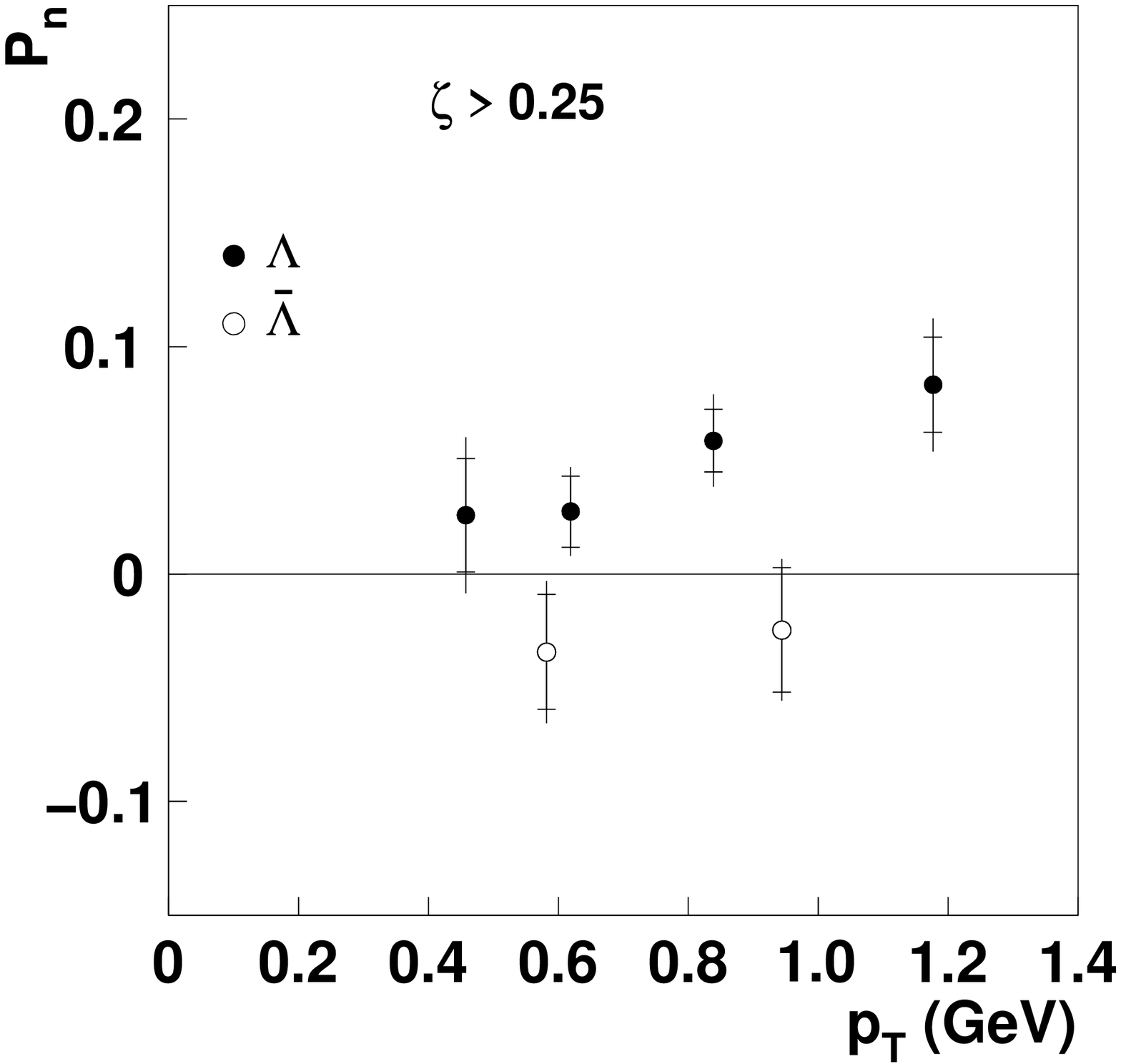} 
  \caption{Transverse polarizations  $\lpol$
        and $\lbarpol$ as a function of $p_{\text{T}}$
        for hyperons from the region $\zeta < 0.25$ (upper panel) and the 
        region $\zeta > 0.25$ (lower panel).
        The inner error bars represent the statistical 
        uncertainties, and the outer error bars represent the statistical and
        systematic uncertainties added in quadrature. 
  }
  \label{fig:PT}
\end{figure}

It should be noted that for each point in $\zeta$ 
the value of the hyperon's mean transverse momentum $\langle p_{\text{T}} \rangle$ 
is different as is shown in the lower panel of Fig.~\ref{fig:zeta}.
Here $p_{\text{T}}$ is defined with respect to the
$eN$ system rather than to the $\gamma^* N$ system as, again, the
virtual-photon direction was not determined in this inclusive
analysis.
In Fig.~\ref{fig:PT}, the transverse $\Lambda$ and $\bar\Lambda$ polarizations 
are shown versus $p_{\text{T}}$ for the two intervals 
$\zeta < 0.25$ and $\zeta > 0.25$.
In both regimes the $\Lambda$ polarization rises linearly with $p_{\text{T}}$,
resembling the linear rise of hyperon polarization 
magnitude with $p_{\text{T}}$
that was consistently observed in the forward production of 
hyperons in hadronic reactions. 
For the $\bar\Lambda$, again no kinematic dependence of the polarization
is observed within statistics.

%

%==========================================================================
% Discussion
%==========================================================================

\section{Discussion}

The transverse $\Lambda$ polarization measured by
HERMES in the $\gamma^* N \rightarrow \Lambda X$ reaction is 
positive, in contrast to
the negative values observed in almost all 
other reactions. 
Very few theoretical models of the kinematic dependence of $\Lambda$
polarization in photo- or electroproduction are available for comparison 
with the data. 
Negative transverse $\Lambda$  and $\bar{\Lambda}$ polarizations were predicted 
for the electroproduction
case in Ref.~\cite{Anselmino2001}, where transverse
$\Lambda$ polarization is associated with the $T$-odd fragmentation function
$D_{1\text{T}}^\perp(z,Q^2)$, one of eight fragmentation functions identified
in a complete tree-level analysis of semi-inclusive deep-inelastic 
scattering~\cite{Mulders1996}.
However, these calculations are confined to the high-$Q^2$
regime of deep-inelastic scattering.

One may speculate on the reason for the positive $\Lambda$
polarization in $\gamma^* N \rightarrow \Lambda X$.
In the model given in Ref.~\cite{DGM}, for example,
forward-going $\Lambda$ particles produced in proton-proton scattering
are formed from the recombination of a high-momentum spin- and isospin-singlet 
\mbox{$ud$ diquark} from the beam with a strange sea quark from the target 
or the fragmentation process. 
The negative $\Lambda$ polarization then arises from the acceleration 
of the strange quark, via the Thomas precession effect.
Conversely, the positive $\Lambda$ 
polarization observed with $K^-$ and $\Sigma^-$ beams is 
indicative of the
deceleration of strange quarks from the beam. 
The positive polarization observed in the HERMES quasireal
photoproduction data might therefore indicate that the 
$\gamma \rightarrow s \bar{s}$ hadronic component of the photon  
plays a significant role in inclusive $\Lambda$ production.

The different average magnitude of $\lpol$ for $\zeta$ below and above
0.25 and the increase of the ratio of $\Lambda$ to $\bar{\Lambda}$
yields at low values of $\zeta$ might be an indication of
different hyperon formation mechanisms in the ``backward'' and
``forward'' kinematic regions, i.e., recombination of a quark from the
beam with a diquark from  the target in the ``backward'' region, and
with a diquark from a string-break in the ``forward'' region.

The positive transverse polarization of $\Lambda$ hyperons has
indeed been 
explained in a quark-recombination model~\cite{Kubo2005},
in which $u$, $d$ and $s$ quarks from the $\gamma$ beam contribute to the $\Lambda$ 
production and polarization through the recombinations $s + (ud)^0$,
$u + (ds)^{0,1}$ and $d + (us)^{0,1}$, where the upper indices 0 (1) 
correspond to singlet (triplet) diquark configurations. The contributions 
of the latter two recombinations are suppressed due to the higher 
mass of the diquarks containing an $s$ quark.

In the framework of impact-parameter-dependent generalized 
parton distribution functions, it was argued in Ref.~\cite{Burkardt2002}
that $\Lambda$ hyperons produced in the collision of a beam containing
$s$ quarks with a nucleon target have a positive transverse polarization. In this
work, a similar mechanism was also used to explain another $T$-odd observable,
the so-called Sivers 
asymmetry in electroproduction of pions as observed at 
HERMES~\cite{hermessivers}\@.

As no theory is currently able to explain the existing body of 
$\Lambda$ polarization data, all such model-dependent speculations must be 
viewed only as exploratory considerations.
The result presented here, a first measurement
of non-zero transverse polarization in the $\gamma^* N \rightarrow \Lambda X$ 
reactions 
at $Q^2 \approx 0$,
adds an interesting new piece to the long-standing mystery of hyperon 
polarization at high energies.\

%\begin{acknowledgements}
We gratefully acknowledge the DESY management for its support and
the DESY staff and the staffs of the collaborating institutions.
This work was supported by
the Ministry of Education and Science of Armenia;
the FWO-Flanders, Belgium;
the Natural Sciences and Engineering Research Council of Canada;
the INTAS and RTN network ESOP contributions
from  the European Union;
the European Commission IHP program;
the German Bundesministerium f\"ur Bildung und Forschung (BMBF); 
the Deutsche Forschungsgemeinschaft (DFG);
the Italian Istituto Nazionale di Fisica Nucleare (INFN);
Monbusho International Scientific Research Program, JSPS, and Toray
Science Foundation of Japan;
the Dutch Foundation for Fundamenteel Onderzoek der Materie (FOM);
the Russian Academy of Science and the Russian Federal Agency for 
Science and Innovations;
the U.K. Particle Physics and Astronomy Research Council; 
and the U.S. Department of Energy (DOE) and the National Science Foundation (NSF).

%\end{acknowledgements}

%==========================================================================
% Bibliography
%==========================================================================

\end{document}

%% file: authors.tex
% List of Institute Addresses 

\def\groupalberta{\affiliation{Department of Physics, University of Alberta, Edmonton, Alberta T6G 2J1, Canada}}
\def\groupargonne{\affiliation{Physics Division, Argonne National Laboratory, Argonne, Illinois 60439-4843, USA}}
\def\groupbari{\affiliation{Istituto Nazionale di Fisica Nucleare, Sezione di Bari, 70124 Bari, Italy}}
\def\groupcolorado{\affiliation{Nuclear Physics Laboratory, University of Colorado, Boulder, Colorado 80309-0390, USA}}
\def\groupdesy{\affiliation{DESY, 22603 Hamburg, Germany}}
\def\groupzeuthen{\affiliation{DESY, 15738 Zeuthen, Germany}}
\def\groupdubna{\affiliation{Joint Institute for Nuclear Research, 141980 Dubna, Russia}}
\def\grouperlangen{\affiliation{Physikalisches Institut, Universit\"at Erlangen-N\"urnberg, 91058 Erlangen, Germany}}
\def\groupferrara{\affiliation{Istituto Nazionale di Fisica Nucleare, Sezione di Ferrara and Dipartimento di Fisica, Universit\`a di Ferrara, 44100 Ferrara, Italy}}
\def\groupfrascati{\affiliation{Istituto Nazionale di Fisica Nucleare, Laboratori Nazionali di Frascati, 00044 Frascati, Italy}}
\def\groupfreiburg{\affiliation{Fakult\"at f\"ur Physik, Universit\"at Freiburg, 79104 Freiburg, Germany}}
\def\groupgent{\affiliation{Department of Subatomic and Radiation Physics, University of Gent, 9000 Gent, Belgium}}
\def\groupgiessen{\affiliation{Physikalisches Institut, Universit\"at Gie{\ss}en, 35392 Gie{\ss}en, Germany}}
\def\groupglasgow{\affiliation{Department of Physics and Astronomy, University of Glasgow, Glasgow G12 8QQ, United Kingdom}}
\def\groupillinois{\affiliation{Department of Physics, University of Illinois, Urbana, Illinois 61801-3080, USA}}
\def\groupliverpool{\affiliation{Physics Department, University of Liverpool, Liverpool L69 7ZE, United Kingdom}}
\def\groupwisconsin{\affiliation{Department of Physics, University of Wisconsin-Madison, Madison, Wisconsin 53706, USA}}
\def\groupmit{\affiliation{Laboratory for Nuclear Science, Massachusetts Institute of Technology, Cambridge, Massachusetts 02139, USA}}
\def\groupmichigan{\affiliation{Randall Laboratory of Physics, University of Michigan, Ann Arbor, Michigan 48109-1040, USA }}
\def\groupmoscow{\affiliation{Lebedev Physical Institute, 117924 Moscow, Russia}}
\def\groupmunich{\affiliation{Sektion Physik, Universit\"at M\"unchen, 85748 Garching, Germany}}
\def\groupnikhef{\affiliation{Nationaal Instituut voor Kernfysica en Hoge-Energiefysica (NIKHEF), 1009 DB Amsterdam, The Netherlands}}
\def\groupstpetersburg{\affiliation{Petersburg Nuclear Physics Institute, St. Petersburg, Gatchina, 188350 Russia}}
\def\groupprotvino{\affiliation{Institute for High Energy Physics, Protvino, Moscow region, 142281 Russia}}
\def\groupregensburg{\affiliation{Institut f\"ur Theoretische Physik, Universit\"at Regensburg, 93040 Regensburg, Germany}}
\def\grouprome{\affiliation{Istituto Nazionale di Fisica Nucleare, Sezione Roma 1, Gruppo Sanit\`a and Physics Laboratory, Istituto Superiore di Sanit\`a, 00161 Roma, Italy}}
\def\groupsimonfraser{\affiliation{Department of Physics, Simon Fraser University, Burnaby, British Columbia V5A 1S6, Canada}}
\def\grouptriumf{\affiliation{TRIUMF, Vancouver, British Columbia V6T 2A3, Canada}}
\def\grouptokyo{\affiliation{Department of Physics, Tokyo Institute of Technology, Tokyo 152, Japan}}
\def\groupamsterdam{\affiliation{Department of Physics and Astronomy, Vrije Universiteit, 1081 HV Amsterdam, The Netherlands}}
\def\groupwarsaw{\affiliation{Andrzej Soltan Institute for Nuclear Studies, 00-689 Warsaw, Poland}}
\def\groupyerevan{\affiliation{Yerevan Physics Institute, 375036 Yerevan, Armenia}}
\def\groupnone{\noaffiliation}

% Set Institute Order 

\groupalberta
\groupargonne
\groupbari
\groupcolorado
\groupdesy
\groupzeuthen
\groupdubna
\grouperlangen
\groupferrara
\groupfrascati
\groupfreiburg
\groupgent
\groupgiessen
\groupglasgow
\groupillinois
\groupliverpool
\groupwisconsin
\groupmit
\groupmichigan
\groupmoscow
\groupmunich
\groupnikhef
\groupstpetersburg
\groupprotvino
\groupregensburg
\grouprome
\groupsimonfraser
\grouptriumf
\grouptokyo
\groupamsterdam
\groupwarsaw
\groupyerevan

% List of Authors 

\author{A.~Airapetian}  \groupyerevan
\author{N.~Akopov}  \groupyerevan
\author{Z.~Akopov}  \groupyerevan
\author{M.~Amarian}  \grouprome \groupyerevan
\author{V.V.~Ammosov}  \groupprotvino
\author{A.~Andrus}  \groupillinois
\author{E.C.~Aschenauer}  \groupzeuthen
\author{W.~Augustyniak}  \groupwarsaw
\author{R.~Avakian}  \groupyerevan
\author{A.~Avetissian}  \groupyerevan
% ...

\collaboration{The HERMES Collaboration} \noaffiliation